\documentclass[preprint,12pt]{elsarticle}

\usepackage{graphicx}

\usepackage{amssymb}

\journal{Physics Letters B}

\begin{document}

\begin{frontmatter}



\title{Ratio of bulk to shear viscosity in a quasigluon plasma:
from weak to strong coupling}


\author[label1,label2]{M.~Bluhm\corref{cor}}
\cortext[cor]{Corresponding author.}
\ead{bluhm@subatech.in2p3.fr}
\author[label3,label4]{B.~K\"ampfer}
\author[label2,label5]{K.~Redlich}

\address[label1]{SUBATECH, UMR 6457, Universit\'{e} de Nantes,
Ecole des Mines de Nantes, IN2P3/CNRS. 4 rue Alfred Kastler,
44307 Nantes cedex 3, France}
\address[label2]{CERN, Physics Department, Theory Devision,
CH-1211 Geneva 23, Switzerland}
\address[label3]{Institut f\"ur Strahlenphysik,
Helmholtz-Zentrum Dresden-Rossendorf, PF 510119, 01314 Dresden, Germany}
\address[label4]{Institut f\"ur Theoretische Physik, TU Dresden,
01062 Dresden, Germany}
\address[label5]{Institute of Theoretical Physics, University of Wroclaw,
PL-50204 Wroclaw, Poland}

\begin{abstract}
The ratio of bulk to shear viscosity is expected to exhibit a different
behaviour in weakly and in strongly coupled systems. This can be expressed
by its dependence on the squared sound velocity.
In the high temperature QCD plasma at small running coupling, the
viscosity ratio is uniquely determined by a quadratic dependence on the
conformality measure, whereas in certain strongly coupled and nearly
conformal theories this dependence is linear. Employing an effective
kinetic theory of quasiparticle excitations with medium-modified dispersion
relation, we analyze the ratio of bulk to shear viscosity of the gluon
plasma. We show that in this approach, depending on the temperature, the
viscosity ratio exhibits either of these dependencies found
by means of weak coupling perturbative or strong coupling holographic techniques.
The turning point between the two different
dependencies is located around the maximum in the scaled interaction measure.
\end{abstract}


\end{frontmatter}


\section{Introduction \label{sec:1}}

The observation that the quark-gluon plasma (QGP) formed in ultra-relativistic
heavy-ion collisions at the Relativistic Heavy Ion Collider (RHIC)
behaves almost like a perfect fluid~\cite{perfect_QGP,perfect_QGP_th},
inspired numerous theoretical efforts to quantify the transport properties
of the produced strongly interacting medium. In particular, various attempts
have been proposed to calculate bulk and shear viscosities for QCD matter by
means of different techniques (see~\cite{Schafer09} for a recent review).
Likewise endeavours have been made to reliably extract information on the
viscosity coefficients from RHIC measurements by continuously improving the
phenomenological analysis of data (see~\cite{Romatschke} for a recent
review). While currently heavy-ion collision phenomenology seems to focus on
the shear viscosity as a relevant quantity, a firm knowledge of the bulk
viscosity as well as other transport coefficients is equally needed to
characterize the properties of the QGP concisely.

The bulk viscosity $\zeta$ is a particular transport coefficient vanishing,
within many models, in the non-relativistic and ultra-relativistic limits as
well as for conformally invariant systems. In strongly interacting matter,
the bulk viscosity to entropy density ratio $\zeta/s$ is expected to be small
outside the deconfinement and chiral phase transition region. In the vicinity
of the transition region, however, it is expected to become large and even to
diverge at a second order phase transition~\cite{Kharzeev08,Karsch08,Romatschke09,
Paech06,Moore08,Sasaki09}. Therefore, a precise knowledge of $\zeta$
might be essential to understand and quantify the transport dynamics of the
strongly interacting fluid in the critical region.

The shear viscosity to entropy density ratio $\eta/s$ for QCD matter, on the
other hand, is expected to be large, i.e. $\eta/s>1$, at large temperatures $T$,
whereas in the vicinity of the deconfinement transition temperature
$T_c$ it exhibits a minimum. Such
a behaviour is known for a variety of liquids and gases~\cite{Kovtun05,Csernai06}
and was conjectured~\cite{Csernai06,Lacey07} to appear also in strongly
interacting matter.

The ratio $\eta/s$ was shown to exhibit a universal behaviour in gauge
theory plasmas at infinite 't Hooft coupling and infinite number
of colours~\cite{Kovtun05,Holo1}, which led to the conjecture of a
fundamental lower bound $(\eta/s)_{KSS}\ge 1/(4\pi)$ (Kovtun-Son-Starinets bound
in natural units) for any physical system~\cite{Kovtun05,Kovtun03}. A similar
universal behaviour for the bulk viscosity or other transport coefficients is
in contrast
not known~\cite{Benincasa06,Buchel06}. Nonetheless, in specific strongly coupled
and nearly conformal theories that allow for a holographically dual
supergravity description~\cite{Buchel05,Buchel08}, the ratio of bulk to shear
viscosity was found to behave as
\begin{equation}
\label{eq1}
 \zeta/\eta \sim \left(\frac13-v_s^2\right) \,,
\end{equation}
where the proportionality constant is of order $\mathcal{O}(1)$ and
$v_s^2=\partial P/\partial \epsilon$ is the squared sound velocity expressed as
derivative of the thermal pressure $P$ with respect to the energy density
$\epsilon$. Recently, a lower bound on this ratio,
$(\zeta/\eta)_B\ge 2\left(1/k-v_s^2\right)$ (Buchel bound), was conjectured for
strongly coupled gauge theories in $k$ spatial dimensions as being valid for all
temperatures, where the holographically dual supergravity description is
allowed~\cite{Buchel08}. This bound is exactly saturated as kinematical
identity for all strongly coupled theories with holographic dual related to
non-conformal branes~\cite{Kanitscheider09}.

In contrast, in scalar field theory~\cite{Horsley87} as well as for photons
interacting with massive particles of a thermal fluid~\cite{Weinberg71}, it was
shown that the ratio $\zeta/\eta$ is uniquely determined by
\begin{equation}
\label{eq2}
 \zeta/\eta = 15\left(\frac13-v_s^2\right)^2 \,.
\end{equation}
In QCD, bulk and shear viscosities were calculated for large temperatures and
small running coupling within kinetic theory~\cite{Arnold06,Arnold}. Under the
relaxation time approximation by assuming equal collision rates for bulk and
shear viscosities~\cite{Arnold06}, these results give rise to a ratio
$\zeta/\eta$ that behaves parametrically like in Eq.~(\ref{eq2}).
Thus, in perturbative QCD (pQCD) the bulk to shear viscosity ratio depends
quadratically on the conformality measure $\Delta v_s^2 = (1/3-v_s^2)$.
This implies that the Buchel bound is violated in high temperature QCD at
small running coupling.

Even though the precise holographic dual to QCD is currently unknown, one could
expect from Eqs.~(\ref{eq1}) and~(\ref{eq2}) that for deconfined strongly
interacting matter the ratio $\zeta/\eta$ undergoes a gradual change from the
non-perturbative to the perturbative regime, rendering its dependence
on the conformality measure from a linear to a quadratic one. It
is therefore interesting to study the temperature dependence of the viscosity
ratio from $T_c$ towards large $T$
in order to understand how the transport properties of the medium are
influenced by its thermal properties.

In this work, considering the gluon plasma as composed of quasiparticle
excitations with medium-modified dispersion relation, we analyze the properties
of the bulk to shear viscosity ratio. The equilibrium thermodynamics of such a
quasiparticle model (QPM) was shown to describe successfully lattice QCD
results on the equation of state and related quantities for pure $SU_c(3)$
gauge theory~\cite{QPM,QPMn}. To study the transport properties of the quasigluon
plasma, we apply an effective kinetic theory approach~\cite{Jeon95+96}. From the
results for bulk and shear viscosities obtained in this way, we show that in a
quasiparticle picture the ratio $\zeta/\eta$ exhibits indeed the above discussed
behaviour with $\Delta v_s^2$: At large temperatures, i.e.~in the
perturbative regime, the dominant dependence of $\zeta/\eta$ on the conformality
measure is quadratic, whereas near the deconfinement transition it renders into
a linear behaviour. The turning point between the two is located in the vicinity
of the maximum in the scaled interaction measure.
Thus, on the level of transport properties, a phenomenological
quasiparticle model of the gluon plasma provides a systematic interpolation between
the regimes of weak and strong coupling.

\section{Effective Kinetic Theory \label{sec:2}}

Assuming that the kinetics of quasigluons is appropriately described by an
effective kinetic equation of Boltzmann-Vlasov type, the phase-space behaviour
of the single-particle distribution function $f$ follows from
$({\cal L} + {\cal V}) f = {\cal C} [f]$, where ${\cal L}$ is the
Liouville operator, ${\cal V}$ is the Vlasov mean field term and
${\cal C}$ is the collision term. In local thermal equilibrium, the
functional ${\cal C} = 0$, and the thermodynamics of the QPM should be
recovered~\cite{QPM,Bluhm05}. This is accomplished by
$f \to n(T) = d\left(\exp(E^0/T) - 1 \right)^{-1}$, where $n(T)$ is the Bose
distribution function for gluon excitations with energy
$E^0 = \sqrt{\vec{p}^{\,\,2} + \Pi(T)}$ and $\Pi(T)$ is the temperature dependent
(but momentum independent) gluon self-energy. The degeneracy factor for colour and
polarization degrees of freedom is $d=16$. Self-consistency of this approach
dictates the form of the Vlasov term. In particular, this implies that in local
thermal equilibrium ${\cal V}$ must be related to the temperature dependence of the
self-energy in order to maintain thermodynamic
self-consistency~\cite{BluhmNPA,BluhmNew}.

\section{Ratio of Bulk to Shear Viscosity \label{sec:3}}

The viscosity coefficients can be derived from the above effective kinetic theory
by assuming small deviations from local thermal equilibrium. The expressions for
bulk and shear viscosities obtained within relaxation time approximation
read~\cite{BluhmNPA,BluhmNew,Chakraborty10}
\begin{eqnarray}
\label{equ:eta2}
 \eta(T) & = & \frac{1}{15T} \int \frac{d^3 \vec{p}}{(2\pi)^3}
 n(T)[1+d^{-1}n(T)] \frac{\tau}{(E^0)^2}\vec{p}^{\,\,4} , \\
\nonumber
 \zeta(T) & = & \frac{1}{T} \int \frac{d^3 \vec{p}}{(2\pi)^3}
 n(T)[1+d^{-1}n(T)] \frac{\tau}{(E^0)^2} \\
\label{equ:zeta3}
 & & \hspace{1.5cm} \times
 \left\{ \left[(E^0)^2-a\right]v_s^2(T)
 - \frac13 \vec{p}^{\,\,2} \right\}^2 ,
\end{eqnarray}
where $\tau$ denotes the relaxation time, which is related to the
collision term ${\cal C}$, and $a = T^2(\partial\Pi(T)/\partial T^2)$.

To quantify the transport coefficients for the quasigluon plasma from
Eqs.~(\ref{equ:eta2}) and~(\ref{equ:zeta3}), we use for the gluon
self-energy~\cite{QPM}
\begin{equation}
\label{equ:selfenergy}
\Pi(T) = \frac12 T^2 G^2(T) \,,
\end{equation}
where $G^2(T)$ is the temperature dependent effective coupling,
\begin{equation}
G^2(T) = \frac{16\pi^2}{11\ln\left[\lambda(T-T_s)/T_c\right]^2} \,,
\end{equation}
defined such that at high temperatures the expressions for bulk
thermodynamic quantities of $SU_c(3)$ gauge theory are recovered by the QPM.
With the parameters $T_s/T_c = 0.73$ and $\lambda = 4.3$,
cf.~\cite{BluhmNew}, pure $SU_c(3)$ lattice gauge theory results for the equation
of state~\cite{Boyd,Okamoto} are nicely described. The quality of this description
is quantified by $\chi^2/$d.o.f.$\,\,= 1.4\cdot 10^{-2}$ for the lattice data
from~\cite{Boyd} and by $\chi^2/$d.o.f.$\,\,= 1.5\cdot 10^{-3}$ for the lattice
data from~\cite{Okamoto}. Within this model, the leading terms of the squared sound
velocity at asymptotically high temperatures, where
$1 \gg G^2 \gg \vert T(dG^2/dT)\vert$ holds, read~\cite{Arnold06}
\begin{equation}
\label{equ:soundspeednew}
v_s^2(T) = \frac13 + \frac{5}{48\pi^2}T\frac{dG^2(T)}{dT} +
\mathcal{O}\left(G^2(T)T\frac{dG^2(T)}{dT}\right) \,,
\end{equation}
which implies that $v_s^2$ reaches $1/3$ only for asymptotically large $T$.

From Eqs.~(\ref{equ:eta2}) and~(\ref{equ:zeta3}) and by assuming an
equal relaxation time for both transport coefficients, the ratio
of bulk to shear viscosity in a quasiparticle description of the gluon
plasma may be written in the form
\begin{eqnarray}
\nonumber
 \frac{\zeta}{\eta} & = & 15\left(\Delta v_s^2\right)^2
 -30\,\Delta v_s^2\,[\Pi(T)-a] \,
 v_s^2(T) \frac{\mathcal{I}_0(T)}{\mathcal{I}_{-2}(T)} \\
\label{equ:zetaetaratio}
 & & \hspace{1.5cm}
 +15[\Pi(T)-a]^2\left(v_s^2(T)\right)^2
 \frac{\mathcal{I}_2(T)}{\mathcal{I}_{-2}(T)} \,,
\end{eqnarray}
with momentum integrals
\begin{equation}
\label{equ:Ik}
 \mathcal{I}_k(T) = \int \frac{d^3 \vec{p}}{(2\pi)^3}
 n(T)[1+d^{-1}n(T)] \frac{\tau}{(E^0)^2}\vec{p}^{\,2-k} \,.
\end{equation}
The momentum integrals receive dominant contributions from thermal
quasiparticle momenta, $|\vec{p}\,|\sim T$, with proportionality
constant of order ${\cal O}(1)$. They follow, for all $T\geq T_c$, the
hierarchy $\mathcal{I}_{-2}/T^2\gg\mathcal{I}_0\gg T^2\mathcal{I}_2>0$.
The terms in Eq.~(\ref{equ:zetaetaratio}), which are proportional to
$[\Pi(T)-a]$, contain only the temperature derivative of the effective
coupling $G^2(T)$, since the leading $T$-dependence cancels out in
this combination. They would, therefore, not be present in the case of
a temperature independent coupling ${g}$, i.e.~in models with a
quasiparticle mass $M(T)$ with trivial temperature dependence of the
form $M(T) \sim T{g}$. Moreover, for constant $M$ one finds
Eq.~(\ref{equ:zetaetaratio}) for the ratio $\zeta/\eta$, but with
$[\Pi(T)-a]$ replaced by $M^2$ and $\Delta v_s^2=0$~\cite{QPM,BluhmNew}.

\begin{figure}[t]
\centering
\includegraphics[scale=0.33]{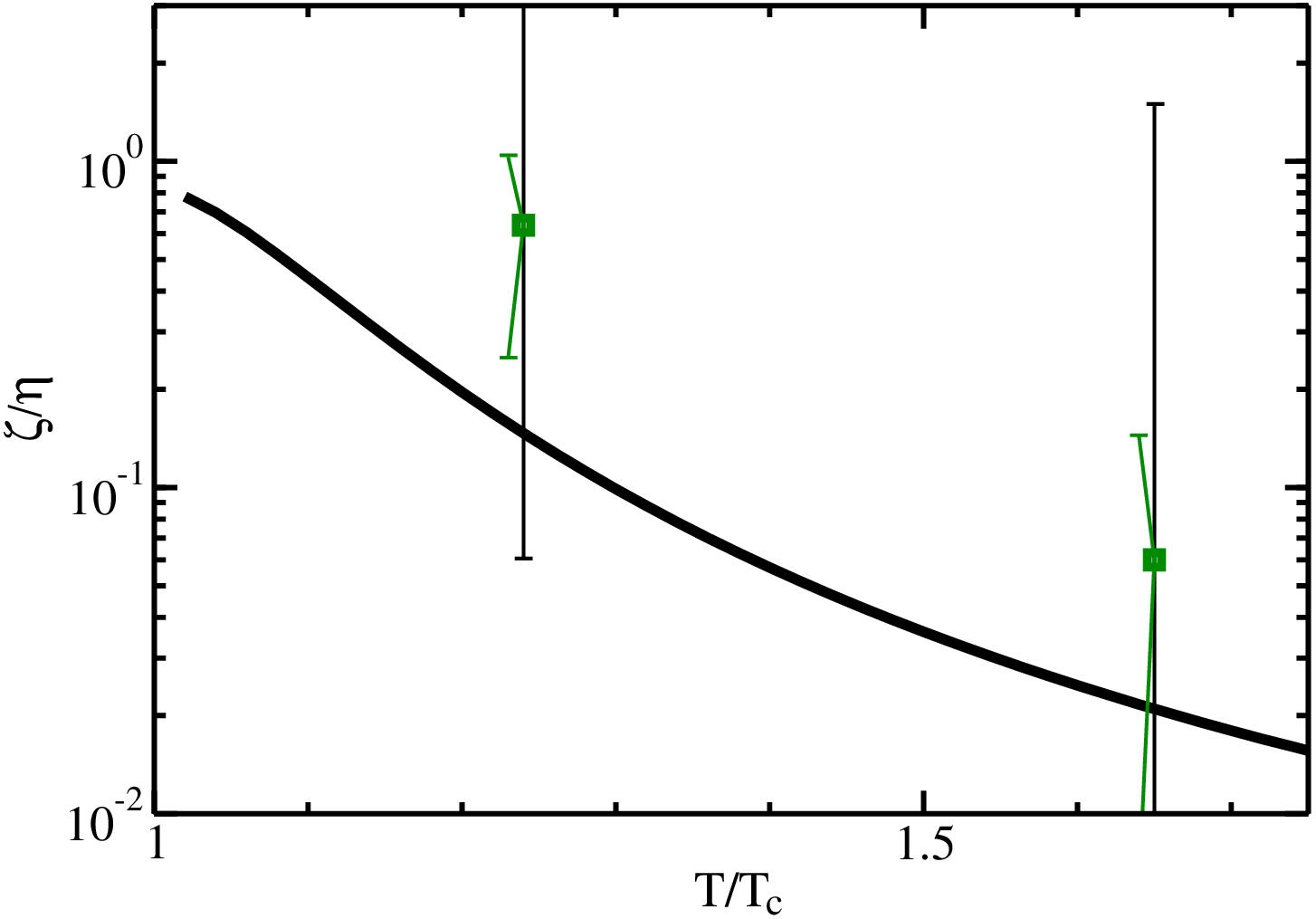}
\caption[]{\label{fig:ratio}
Scaled temperature dependence of the bulk to shear viscosity ratio
Eq.~(\ref{equ:zetaetaratio}) (solid curve) for
$1.02\leq T/T_c\leq 1.75$ compared with available lattice QCD results
from~\cite{Meyer} (squares). The displaced error bars are determined from the
statistical errors in $\zeta$ and $\eta$~\cite{Meyer}. The
non-displaced error bars are obtained from
the conservative upper and lower bounds on $\zeta$ given in~\cite{Meyer}
and the statistical errors in $\eta$.}
\end{figure}

The ratio $\zeta/\eta$ from Eq.~(\ref{equ:zetaetaratio}) is quantified in
Fig.~\ref{fig:ratio} as a function
of $T/T_c$ for a gluon plasma described by quasiparticle excitations with
medium-modified dispersion relation and a momentum independent relaxation
time $\tau$. For comparison, also shown in Fig.~\ref{fig:ratio} is the bulk
to shear viscosity ratio obtained from lattice QCD results in~\cite{Meyer}.
The bulk and shear viscosities calculated in lattice QCD are fairly well
reproduced within our quasiparticle model~\cite{BluhmNew}.
Noticing the large errors in Fig.~\ref{fig:ratio}, the temperature dependence
of the viscosity ratio in the quasigluon plasma is seen in Fig.~\ref{fig:ratio}
to be as well consistent with that found from the lattice gauge theory
calculations.

For $T\gtrsim 1.5\,T_c$, the viscosity ratio is entirely determined by the
pQCD-like behaviour from Eq.~(\ref{eq2}). However, for $T$ near $T_c$, the
non-perturbative contributions in Eq.~(\ref{equ:zetaetaratio}) become
significant, resulting in a reduction of the $\zeta/\eta$ ratio.
As evident from Fig.~\ref{fig:ratio},
for $T\gtrsim 1.02\,T_c$, the bulk viscosity is smaller than the
shear viscosity in the quasigluon plasma. For $T\rightarrow 1.02\,T_c$, the
ratio $\zeta/\eta$ increases and reaches about $0.78$. Quite similar values
for this ratio were reported in holographic
approaches~\cite{Buchel09,Guersoy09,Guersoy10} and used in hydrodynamic
simulations when studying the influence of a finite $\zeta$ and $\eta$ on
the elliptic flow~\cite{Song10}. Nonetheless, we note that in a different
holographic approach~\cite{Buchel08} a much larger value for $\zeta/\eta$
in this temperature region was predicted.

The viscosity ratio $\zeta/\eta$ in Eq.~(\ref{equ:zetaetaratio}) suggests
apparently both, a quadratic behaviour
$\propto \left(\Delta v_s^2\right)^2$, as
found in the regime of small QCD running coupling~\cite{Arnold06}, and
a linear behaviour $\propto \Delta v_s^2$, which is in line with the
dependence found in specific strongly coupled systems near
conformality~\cite{Benincasa06,Buchel05,Buchel08}. A similar combination
of dependencies was also pointed out in~\cite{Buchel05,Buchel09}
as phenomenologically relevant relation.
One should emphasize, however, that all other quantities entering the $\zeta/\eta$ ratio
in Eq.~(\ref{equ:zetaetaratio}) depend on $T=T(\Delta v_s^2)$. Therefore,
Eq.~(\ref{equ:zetaetaratio}) cannot be understood as a strict expansion
series in powers of the conformality measure. Rather, for a quantification any temperature 
dependence has to be converted into a $\Delta v_s^2$-dependence. 

The behaviour of $T$ with $\Delta v_s^2$ can be obtained 
by inverting the relation between the squared sound velocity $v_s^2(T)$ and 
the temperature. In the considered quasigluon plasma, $v_s^2(T)$ follows as 
\begin{equation}
\label{equ:soundspeed}
 v_s^2(T) = \frac13 \frac{1}{\left[1 + \frac{T}{3}
 \frac{\mathcal{J}_1(T)}{\mathcal{J}_2(T)}
 \frac{d}{dT} \left( \frac{\Pi (T)}{T^2}\right)\right]} ,
\end{equation}
where the entering integrals read 
\begin{eqnarray}
\nonumber
 \mathcal{J}_1 & = & \int_0^\infty dx \frac{x^2}{\left(e^{\sqrt{x^2+z}}-1\right)}\Bigg(
 \frac{1}{\sqrt{x^2+z}}-\frac{\left(\frac43 x^2+z\right)}{2\sqrt{x^2+z}^{\,3}} \\
\nonumber
 & & \hspace{2.5cm} -
 \frac{\left(\frac43 x^2+z\right)e^{\sqrt{x^2+z}}}{2(x^2+z)\left(e^{\sqrt{x^2+z}}-
 1\right)}\Bigg) 
\end{eqnarray}
and 
\begin{eqnarray}
\nonumber
 \mathcal{J}_2 & = & \int_0^\infty dx \frac{x^2\left(\frac43 x^2+
 z\right)}{\left(e^{\sqrt{x^2+z}}-1\right)\sqrt{x^2+z}} 
\end{eqnarray}
with $z=\Pi(T)/T^2$. For asymptotically large $T$, Eq.~(\ref{equ:soundspeed})
simplifies to Eq.~(\ref{equ:soundspeednew}).

\begin{figure}[t]
\centering
\includegraphics[scale=0.27]{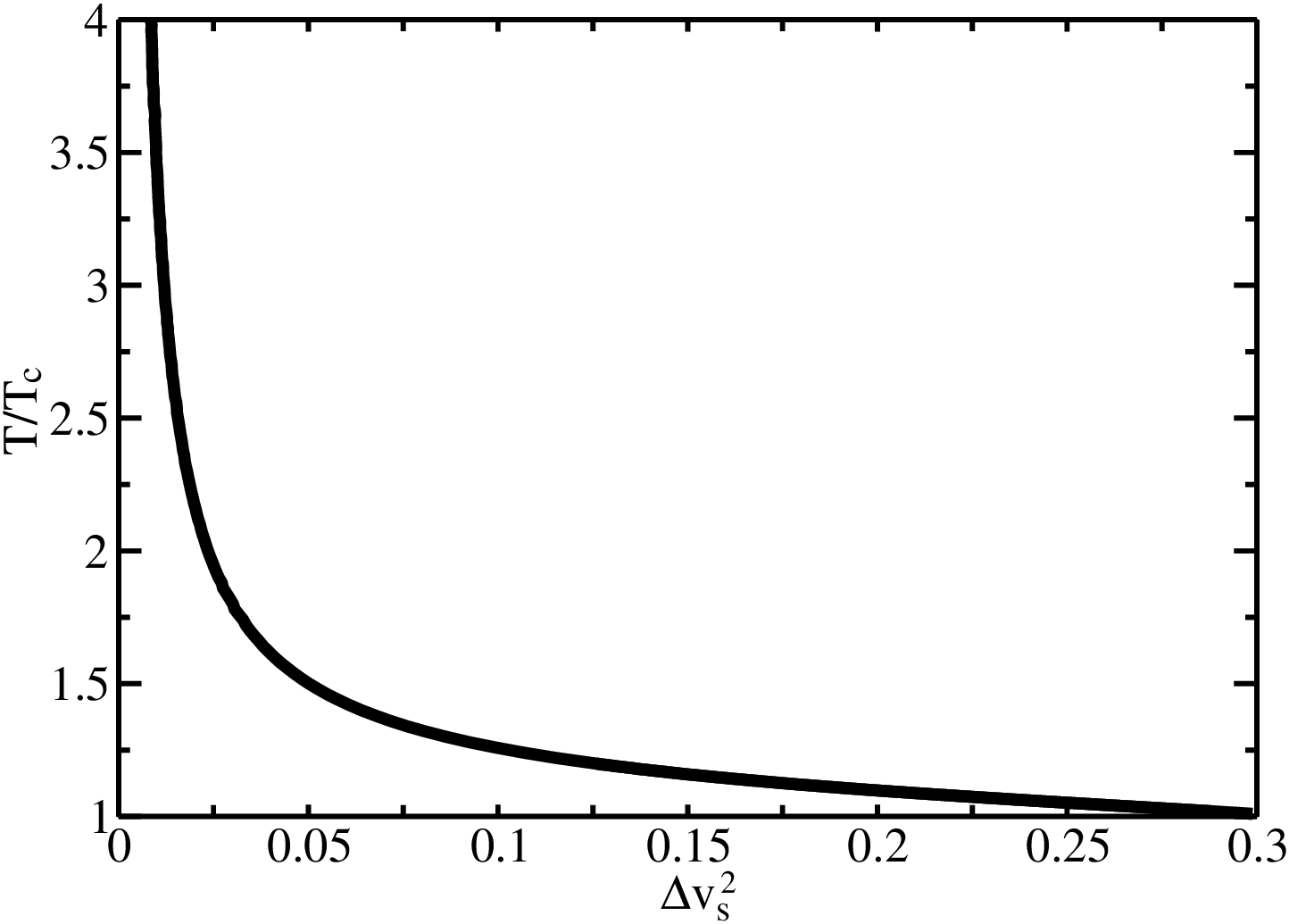}
\hspace{1mm}
\includegraphics[scale=0.27]{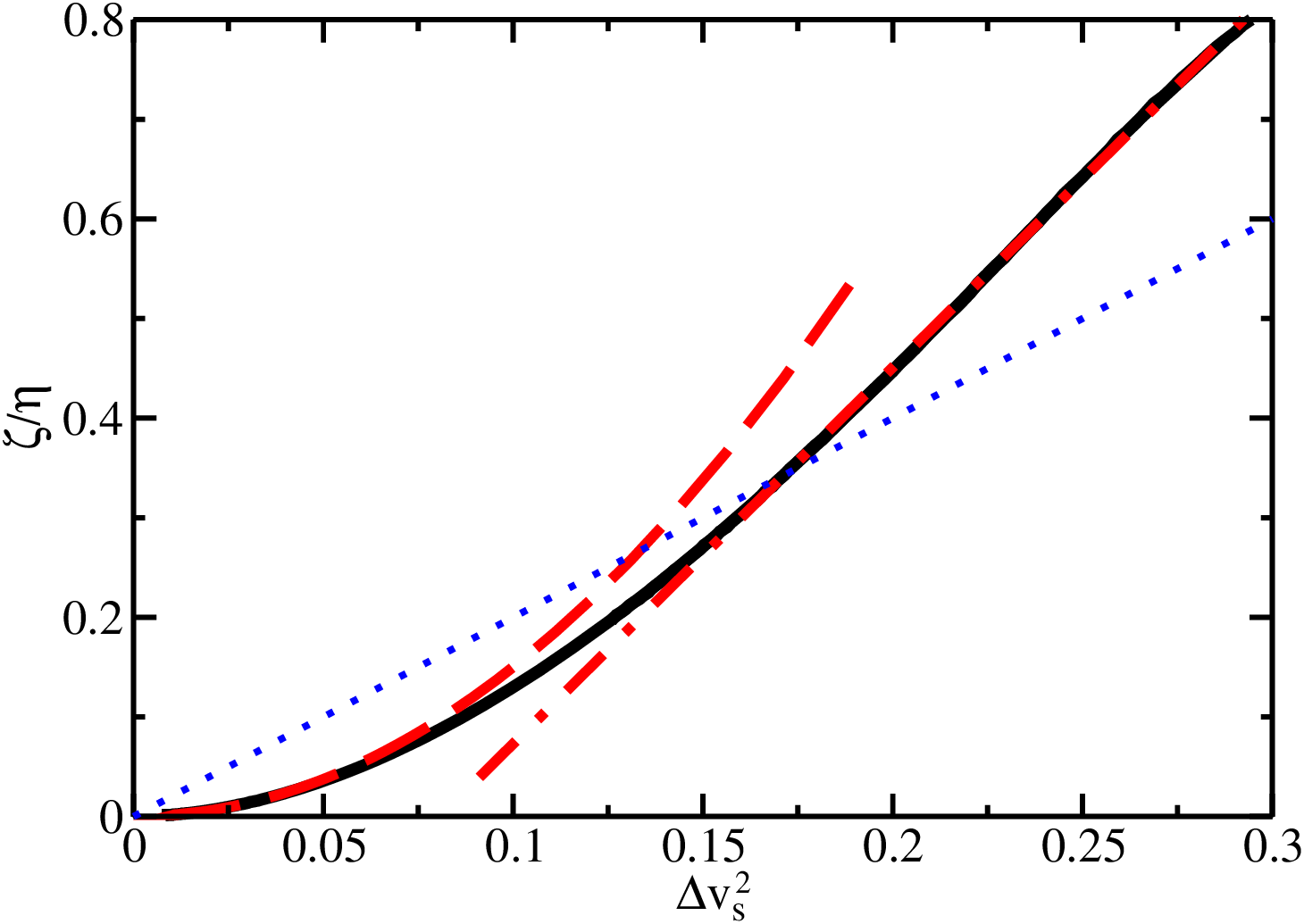}
\caption[]{\label{fig:ratio2}
Left: Scaled temperature dependence of $\Delta v_s^2$ in the QPM.
Right: The ratio $\zeta/\eta$ from Eq.~(\ref{equ:zetaetaratio}) (solid curve)
as a function of the conformality measure $\Delta v_s^2$. The dash-dotted curve
shows a linear fit $\zeta/\eta=\alpha\,\Delta v_s^2 + \beta$ with
$\alpha=3.78$ and $\beta=-0.305$. The dashed curve depicts the quadratic
contribution in Eq.~(\ref{equ:zetaetaratio}), $15\left(\Delta v_s^2\right)^2$.
The dotted curve shows Buchel's conjectured lower bound for
$\zeta/\eta$.}
\end{figure}
The relation $T=T(\Delta v_s^2)$ following from Eq.~(\ref{equ:soundspeed}) 
is shown in the left panel of Fig.~\ref{fig:ratio2}. The right panel of this figure 
quantifies the resulting dependence of the ratio $\zeta/\eta$ on the conformality 
measure $\Delta v_s^2$ in the interval $0.01\leq\Delta v_s^2\leq 0.3$. 
For small values of $\Delta v_s^2$, that is for large temperatures $T\gtrsim 1.5\,T_c$
(see left panel of Fig.~\ref{fig:ratio2}), the viscosity ratio
shows a quadratic behaviour with $\Delta v_s^2$. Indeed,
for asymptotically large $T$, one finds with Eq.~(\ref{equ:soundspeednew}), that
$\Delta v_s^2\simeq -5/(24\pi^2)T^2(dG^2/dT^2)$, where
$\vert T^2(dG^2/dT^2)\vert\ll 1$. Then, all terms in Eq.~(\ref{equ:zetaetaratio})
are at leading order in $G^2$
proportional to $\left(T^2dG^2/dT^2\right)^2$, which implies
in turn, that in this temperature regime all terms in Eq.~(\ref{equ:zetaetaratio})
are at leading order proportional to $\left(\Delta v_s^2\right)^2$. However, since
$T^4\mathcal{I}_2/\mathcal{I}_{-2}\ll T^2\mathcal{I}_0/\mathcal{I}_{-2}\ll 1$,
the numerically dominant contribution to $\zeta/\eta$ in
Eq.~(\ref{equ:zetaetaratio}) is given by $15\left(\Delta v_s^2\right)^2$,
and Eq.~(\ref{eq2}) is recovered.

With increasing
$\Delta v_s^2$, i.e.~for decreasing $T$ (see left panel of Fig.~\ref{fig:ratio2}),
the bulk to shear viscosity ratio renders numerically
into a linear rise with $\Delta v_s^2$ as seen in the right panel
of Fig.~\ref{fig:ratio2}. In fact,
with decreasing $T$ towards $T_c$, the quadratic term $15 \left(\Delta v_s^2\right)^2$ in
Eq.~(\ref{equ:zetaetaratio})
is cancelled to a large degree of accuracy by the other terms,
which in the regime $0.17\lesssim \Delta v_s^2\leq 0.3$ sum up to
approximately $-15 \left(\Delta v_s^2\right)^2 + \alpha\,\Delta v_s^2 + \beta$
with constant coefficients $\alpha$ and $\beta$. Thus, a linear relation
$\zeta/\eta = \alpha\,\Delta v_s^2 + \beta$ emerges as evident from
Fig.~\ref{fig:ratio2} (right panel). The onset of this
change in the behaviour with
the conformality measure is located around $T\simeq 1.15\,T_c$. This lies
in the direct proximity of the maximum in the scaled interaction
measure~\cite{BluhmNew}.

Fig.~\ref{fig:ratio2} (right panel) depicts, in addition, Buchel's lower
bound on the ratio $\zeta/\eta$~\cite{Buchel08} (dotted curve). In our
quasiparticle description of the gluon plasma this bound is satisfied
also only for temperatures $T\lesssim 1.15\,T_c$, when
$\Delta v_s^2\gtrsim 0.17$. We note, however, that this observation is
based on the used approximation of a momentum independent relaxation time,
which is the same for both viscosity coefficients. Therefore, it would
be interesting to study how a momentum dependent $\tau$ in
Eq.~(\ref{equ:zetaetaratio}), as proposed in~\cite{Chakraborty10}, would
influence the quantitative comparison of Eq.~(\ref{equ:zetaetaratio})
with the conjectured lower bound and the quantitative dependence of the
viscosity ratio on $\Delta v_s^2$.

This demonstrates that within an effective kinetic theory approach
for quasigluon excitations, which is supplemented by information from equilibrium
thermodynamics, both limiting behaviours of the bulk to shear viscosity ratio
with $\Delta v_s^2$ known from the weak~\cite{Arnold06} and strong coupling
regimes~\cite{Benincasa06,Buchel05,Buchel08} can be found. Thus, the
quasiparticle model for the gluon plasma provides a phenomenological method
to describe non-perturbative effects. Within such a model, one can effectively
describe thermodynamics as well as transport properties of the gluon plasma in
the whole temperature range from asymptotically large temperatures towards the
critical region just above deconfinement.

\section{Conclusion \label{sec:4}}

Within our quasiparticle model for the gluon plasma, which is based on an
effective kinetic theory, we calculated explicitly the bulk to shear viscosity
ratio $\zeta/\eta$ in the relaxation time approximation. We have shown, that
this ratio exhibits, depending on $T$,
a quadratic or a linear behaviour with the conformality
measure $\Delta v_s^2$ of the medium. Similar dependencies are known from small
running coupling QCD~\cite{Arnold06} and holographic~\cite{Buchel05,Buchel08}
approaches, respectively. While at asymptotically large $T$ the quadratic
dependence on $\Delta v_s^2$ is dominant in our model,
the linear dependence becomes more important
when approaching $T_c$, governing effectively the qualitative behaviour of the viscosity
ratio for $1.02\lesssim T/T_c\lesssim 1.15$.
The turning point in the behaviour is located around the maximum in the scaled interaction
measure. Therefore, besides a proper description
of equilibrium thermodynamics of pure $SU_c(3)$ lattice gauge theory, our quasiparticle
model of the gluon plasma also describes its viscosity coefficients, cf.~\cite{BluhmNew},
and reproduces the dependencies of their ratio on $\Delta v_s^2$ as expected in the
weak and strong coupling regimes, thus, providing a systematic link between the two.
For momentum independent relaxation times, assumed to be common for bulk and shear
viscosities, we find a significant dependence of $\zeta/\eta$ on $v_s^2$.
This might imply consequences for viscous hydrodynamic simulations, cf.~the studies
in~\cite{Song10,Simulations}, in particular for LHC energies, where a larger
temperature range is probed. However, for any phenomenological application one
would have to include quark degrees of freedom into the model calculations.

\section*{Acknowledgements}
Stimulating discussions with A.~Buchel, S.~Jeon, U. G\"ursoy, E. Kiritsis, H.~B.~Meyer,
S.~Peign\'{e}, A.~Peshier, C.~Sasaki and U.~Wiedemann are kindly acknowledged.
We also thank R.~Kotte for his help on error propagation in combined data sets.
This work is supported by BMBF 06DR9059, the European Network I3-HP2 Toric and
the Polish Ministry of Science.
We also acknowledge the CERN TH Institute "The first heavy ion collisions at the LHC",
where this work was advanced.








\end{document}